\documentclass[amsmath,amssymb,aps,letterpaper,prl,twocolumn,longbibliography,10pt]{revtex4-1}
\pdfoutput=1
\usepackage{graphicx}
\usepackage{color}
\usepackage{comment}

\usepackage{subfigure}

\usepackage{color}

\makeatletter \let\@keywords\@empty \let\@subject\@empty
\providecommand{\keywords}[1]{\gdef\@keywords{#1}}
\providecommand{\subject}[1]{\gdef\@subject{#1}}
\def\thetitle{\@title}
\def\theauthor{\@author}
\def\thesubject{\@subject}
\def\thedate{\@date}
\def\thekeywords{\@keywords}
\makeatother
\AtBeginDocument{%
\hypersetup{pdftitle={\thetitle}}%
\hypersetup{pdfauthor={\theauthor}}%
\hypersetup{pdfsubject={\thesubject}}%
\hypersetup{pdfkeywords={\thekeywords}}%
}

\usepackage[bookmarks=true,hyperfigures=true]{hyperref}
\usepackage{fixmath}
\usepackage{mathtools}

\providecommand{\href}[2]{#2}

\let\oldbfseries=\bfseries
\let\oldmdseries=\mdseries
\let\oldnormalfont=\normalfont
\renewcommand{\bfseries}{\oldbfseries\boldmath}
\renewcommand{\mdseries}{\oldmdseries\unboldmath}
\renewcommand{\normalfont}{\oldnormalfont\unboldmath}

\makeatletter
\newlength{\apb@width}
\newcommand{\autoparbox}[2][c]{\settowidth{\apb@width}{#2}\parbox[#1]{\apb@width}{#2}}

\makeatother

\newcommand{\e}{\operatorname{e}}
\DeclareMathOperator{\checkstar}{\check {\star}}
\DeclareMathOperator{\hatstar}{\hat {\star}}
\newcommand{\de}{\operatorname{d}\!}
\DeclareMathOperator{\tr}{tr}
\newcommand{\eqndot}{\, . }
\newcommand{\eqncom}{\, , }

\DeclareMathOperator{\phaneq}{\phantom{{}=}}

\newcommand{\YM}{{\mathrm{\scriptscriptstyle YM}}}
\newcommand{\CY}{\mathcal{Y}}
\newcommand{\noty}{z}
\newcommand{\yy}{\noty}

\renewcommand{\Im}{\operatorname{Im}}

\begin{document}

\title{The Hagedorn temperature of \texorpdfstring{AdS$_5$/CFT$_4$}{AdS5/CFT4} via integrability}

\author{Troels Harmark}%
 \email{harmark@nbi.ku.dk}
\author{Matthias Wilhelm}%
  \email{matthias.wilhelm@nbi.ku.dk}

\affiliation{%
Niels Bohr Institute, Copenhagen University,\\
Blegdamsvej 17, 2100 Copenhagen \O{}, Denmark
}%

\begin{abstract}
We establish a framework for calculating the Hagedorn temperature of AdS$_5$/CFT$_4$ via integrability.
Concretely, we derive the thermodynamic Bethe ansatz equations that yield the Hagedorn temperature of planar $\mathcal{N}=4$ super Yang-Mills theory at any value of the 't Hooft coupling.
We solve these equations perturbatively at weak coupling via the associated Y-system, confirming the known results at tree level and one-loop order as well as deriving the previously unknown two-loop Hagedorn temperature. 
Finally, we comment on solving the equations at finite coupling.
\end{abstract}

\maketitle

\section{Introduction}

According to the AdS/CFT correspondence \cite{Maldacena:1997re}, $\mathcal{N}=4$ super Yang-Mills (SYM) theory on $\mathbb{R} \times S^3$ is dual to type IIB string theory on $\mbox{AdS}_5\times S^5$. This duality should in particular relate the phase transitions, critical behavior and thermal physics of the theories.  

One interesting example of a critical behavior is the Hagedorn temperature. In the planar limit of $\mathcal{N}=4$ SYM theory on $\mathbb{R} \times S^3$, the origin of the Hagedorn temperature $T_{\text{H}}$ is the confinement of the color degrees of freedom due to the theory being on a three-sphere. This enables the theory to have a phase transition that bears resemblance to the confinement/deconfinement phase transition in QCD or pure Yang-Mills theory \cite{Atick:1988si,Aharony:2003sx}.

The Hagedorn temperature is the lowest temperature for which the planar partition function $\mathcal{Z}(T)$
diverges. Via the state/operator correspondence, the partition function can be re-expressed in terms of the dilatation operator $D$ of $\mathcal{N}=4$ SYM theory on $\mathbb{R}^4$:
\begin{equation}
 \mathcal{Z}(T)=\tr_{\mathbb{R}\times S^3}[\e^{-H/T}]=\tr_{\mathbb{R}^4}[\e^{-D/T}]\eqncom
\end{equation}
where we have set the radius of $S^3$ to $1$.
States correspond to gauge-invariant operators consisting of one or more trace factors.
The energies correspond to the scaling dimensions of the operators, as measured by the dilatation operator.
In the planar limit, the scaling dimensions of multi-trace operators are entirely determined by those of their single-trace factors, and the latter can be enumerated via P\'{o}lya theory to determine the partition function and thus the Hagedorn temperature in the free theory \cite{Sundborg:1999ue}.
This procedure was later generalized to one-loop order and to the case of non-zero chemical potentials \cite{Spradlin:2004pp,Yamada:2006rx,Harmark:2006di,Suzuki:2017ipd,GomezReino:2005bq}.

On the string-theory side, the Hagedorn temperature occurs due to the exponential growth of string states with the energy present in tree-level string theory. For interacting string theory, it is connected to the Hawking-Page phase transition \cite{Witten:1998zw}. This suggests that the confinement/deconfinement transition on the gauge-theory side is mapped on the string-theory side to a transition from a gas of gravitons (closed strings) for low temperatures to a black hole for high temperatures.
In particular, also the Hagedorn temperature on the gauge-theory and string-theory sides of the AdS/CFT correspondence should be connected 
 \cite{Sundborg:1999ue,Aharony:2003sx}. 
 
On the string-theory side, the Hagedorn temperature has been computed in pp-wave limits \cite{PandoZayas:2002hh,Greene:2002cd,Brower:2002zx,Grignani:2003cs}. In \cite{Harmark:2006ta}, the first quantitative interpolation of the Hagedorn temperature from the gauge-theory side to the string-theory side was made, exploiting a limit towards a critical point in the grand canonical ensemble \cite{Harmark:2014mpa}.
This limit effectively reduces the gauge-theory side to the $\mathfrak{su}(2)$ sector with only the one-loop dilatation operator surviving, which enables one to match the Hagedorn temperature of the gauge-theory side to that of string theory on a pp-wave background via the continuum limit of the free energy of the Heisenberg spin chain. 

A hitherto unrelated but very powerful property of planar $\mathcal{N}=4$ SYM theory is integrability, see \cite{Beisert:2010jr,Bombardelli:2016rwb} for reviews. It amounts to the existence of an underlying two-dimensional exactly solvable model, which reduces to an integrable sigma model at strong coupling and to an integrable spin chain at weak coupling. 
Via integrability, the planar scaling dimensions of all single-trace operators can in principle be calculated at any value of the 't Hooft coupling $\lambda=g_\YM^2N$, allowing for a smooth interpolation between weak and strong coupling results.
In practice, however, the calculation for each operator is so involved that summing the results for all operators to obtain the partition function $\mathcal{Z}(T)$ seems prohibitive.

In this letter, we show how to use integrability to compute the Hagedorn temperature at any value of the 't Hooft coupling.
In the spectral problem, the integrable model is solved 
on a cylinder of finite circumference $L$, which accounts for wrapping contributions to the scaling dimension due to the finite length of the spin chain.
In order to calculate the partition function $\mathcal{Z}(T)$, we would need to solve this model on the torus with circumferences $L$ and $1/T$, an endeavor that has not been successful yet even for the Heisenberg spin chain.
The Hagedorn singularity, however, is driven by the contributions of spin chains with very high $L$, or rather very high classical scaling dimension, 
where the finite-size corrections play no role \footnote{The fact that finite-size corrections are suppressed in $L$ is manifest. The suppression in the classical scaling dimension can for example be seen for so-called twist operators. Up to corrections in the classical scaling dimension, these operators are dual to Wilson loops, for which no finite-size effects occur \cite{Korchemsky:1988si,Belitsky:2003ys}.}. 
Thus, we can calculate it by solving the integrable model on a cylinder of circumference $1/T$, a situation that is related to the one in the spectral problem via a double Wick rotation.
Indeed, we find a direct relation between the continuum limit of the free energy of the spin chain associated with planar $\mathcal{N}=4$ SYM theory and the Hagedorn temperature.
Using the integrability of the model, we derive thermodynamic Bethe ansatz (TBA) equations which determine the Hagedorn temperature at any value of the 't Hooft coupling. We present them in the form of a Y-system in \eqref{eq: general Y system}--\eqref{eq: theta n}.
As a first application, we solve them in the constant case as well as perturbatively at weak coupling, confirming the known tree-level and one-loop Hagedorn temperature. Moreover, we determine the previously unknown two-loop Hagedorn temperature:
\begin{equation}
\begin{aligned}
 T_{\text{H}}&=\frac{1}{2\log(2+\sqrt{3})}+\frac{1}{\log(2+\sqrt{3})}g^2 \\
 &\phaneq+
 \left( -\frac{86}{ \sqrt{3}} + \frac{24 \log(12)}{\log(2+\sqrt{3})}\right)
 g^4+\mathcal{O}(g^6)\,,
 \end{aligned}
\end{equation}
where $g^2=\frac{\lambda}{16\pi^2}$.

\section{TBA equations for the Hagedorn temperature}

In the following, we relate the Hagedorn temperature to the spin-chain free energy and derive TBA equations for the latter. 

\paragraph{The Hagedorn temperature from the free energy of the spin chain}

In the planar limit, the scaling dimensions of multi-trace operators are completely determined by the scaling dimensions of their single-trace factors. The partition function $\mathcal{Z}(T)$ is then entirely determined by the single-trace partition function $Z(T)$.
Splitting the dilatation operator into a classical and an anomalous part as $D=D_0+\delta D$, 
we can write 
\begin{equation}
 Z(T)=\sum_{m=2}^\infty\e^{-\frac{m}{2}\frac{1}{T}(1 + F_m(T))}\eqncom
\end{equation}
where 
\begin{equation}
\label{eq: spin_chain_Z}
 F_m(T)=-T\frac{2}{m}\log \left( \tr_{\text{spin-chain},D_0=\frac{m}{2}}[\e^{-\delta D/T}] \right)    
 \end{equation}
is the spin-chain free energy per unit classical scaling dimension for fixed $D_0=\frac{m}{2}$. 
The multi-trace partition function $\mathcal{Z}(T)$ is then given by 
\begin{equation}
\label{eq: relation between partition function and free energy}
 \mathcal{Z}(T)=\exp\sum_{n=1}^\infty\frac{1}{n}\sum_{m=2}^\infty(-1)^{m(n+1)}\e^{-\frac{m}{2T} (n + F_m(T/n))}\eqncom
\end{equation}
where the alternating sign takes care of the correct statistics.
The Hagedorn singularity is the first singularity of $\mathcal{Z}(T)$ encountered raising the temperature from zero.
It arises from the $n=1$ contribution to the sum over $n$, i.e.\ from the infinite series
\begin{equation}
\label{n1contr}
\sum_{m=2}^\infty \e^{-\frac{m}{2T} (1 + F_m(T))}\eqncom
\end{equation}
where each term in the series is finite as $F_m(T)$ only includes a finite number of states.
We can use Cauchy's root test to assess when this series diverges. To this end, we compute the $m$th root of the absolute value of the $m$th term and take the large $m$ limit, giving
\begin{equation}
r = \lim_{m\rightarrow \infty} \e^{-\frac{1}{2T}(1+F_m(T))} = \e^{-\frac{1}{2T} ( 1+ F(T))} \eqncom
\end{equation}
where
\begin{equation}
 F(T)= \lim_{m\rightarrow \infty} F_m(T)
\end{equation}
is the thermodynamic limit of the free energy.
The root test states that the series is convergent for $r<1$ and divergent for $r>1$. Thus, the Hagedorn temperature is  determined from $r=1$ or, equivalently, from
\begin{equation}
\label{eq: hagedorn temperature via F}
F(T_{\text{H}})=-1\eqndot
\end{equation}

\paragraph{TBA equations}

The free energy $F$ of the spin chain can be calculated via the thermodynamic Bethe ansatz (TBA). The TBA equations for the Hagedorn temperature of $\mathcal{N}=4$ SYM theory can be derived in analogy to the case of the spectral problem \cite{Arutyunov:2009zu,Bombardelli:2009ns,Gromov:2009bc,Arutyunov:2009ur,Gromov:2009tv,Cavaglia:2010nm}.
The starting point are the all-loop asymptotic Bethe equations \cite{Beisert:2004hm,Beisert:2006ez} for the 
$\mathfrak{psu}(2,2|4)$ spin chain found in the spectral problem,
which are written in terms of the length $L$ of the spin chain as well as the seven excitation numbers corresponding to the roots in the Dynkin diagram of $\mathfrak{psu}(2,2|4)$. We then rewrite the Bethe equation so that the middle, momentum-carrying root is written in terms of $D_0$ instead of $L$, since it is $D_0$ that we keep fixed when calculating the free energy \eqref{eq: spin_chain_Z}. We proceed by employing the string hypothesis, which  enables us to write the Bethe equations for many magnons. The next step is the continuum limit $D_0 \rightarrow \infty$, in which we can write the TBA equations in terms of the Y-functions defined from the densities of the strings. In particular, it allows us to write down the free energy. 
The main difference compared to the TBA equations of the spectral problem is that we do not make a double Wick rotation i.e.\ we consider the so-called direct theory and not the mirror theory. This means we use the Zhukovsky variable $x(u)$ with a short cut: 
\begin{equation}
x(u)=\frac{u}{2}\left(1+\sqrt{1-\frac{4g^2}{u^2}}\right)\eqndot
\end{equation}
Note that TBA equations for the direct theory were also considered in \cite{Cavaglia:2010nm,Arutynov:2014ota} but in different thermodynamic limits.

\paragraph{Y-system}
The TBA equations can be rephrased in terms of a Y-system consisting of the functions $\CY_{a,s}$, where $(a,s)\in M=\{(a,s)\in\mathbb{N}_{\geq0}\times\mathbb{N} \,|\,a=1 \vee |s|\leq 2 \vee \pm s=a=2\}$.
With some exceptions, they satisfy the  equations
\begin{equation}
\label{eq: general Y system}
 \log\CY_{a,s}=\log\frac{(1+\CY_{a,s-1})(1+\CY_{a,s+1})}{(1+\CY_{a-1,s}^{-1})(1+\CY_{a+1,s}^{-1})}\star s\eqncom
\end{equation}
where $\star$ denotes the convolution with $s(u)=(2 \cosh \pi u)^{-1}$ on $\mathbb{R}$ and the (inverse) Y-functions with shifted indices are assumed to be zero when the shifted indices are not in $M$.
The Y-functions are analytic in the strip with $|\Im(u)|<\frac{1}{2}|a-|s||$.
For the purpose of this letter, the chemical potentials are set to zero. Hence, the Y-system is symmetric, $\CY_{a,s}=\CY_{a,-s}$, with boundary conditions 
\begin{equation}
\label{eq: Ybcs}
\lim_{a\rightarrow \infty} \frac{\CY_{a+1,s}}{\CY_{a,s}} = 1 \eqncom \quad  \lim_{n\rightarrow \infty} \frac{\CY_{1,n+1}}{\CY_{1,n}} = 1 \eqncom
\end{equation}
for $s=0,\pm 1$.
The first of the aforementioned exceptions to the  equations \eqref{eq: general Y system} then is 
\begin{equation}
\label{eq: CY_10}
 \log \CY_{1,0}  = - \rho\hatstar s  + 2\log (1+\CY_{1,1})\checkstar s-\log(1+\CY_{2,0}^{-1})\star s\,,
\end{equation}
where we have defined $\hatstar$ and $\checkstar$ as the convolutions on $(-2g,2g)$ and $\mathbb{R}\setminus(-2g,2g)$, respectively.
Similarly, the convolution with $\CY_{1,1}$ and $\CY_{2,2}$ in \eqref{eq: general Y system} for $(a,s)=(2,1),(1,2)$ is also understood to be $\checkstar$.
The source term $\rho(u)$ is defined as 
\begin{equation}
 \begin{aligned}
\rho &=  \frac{\epsilon_0}{T}    + 2 \log (1+\CY_{1,1}) (1+\CY_{2,2}^{-1}) \checkstar  H_0   \\ &\phaneq  + 2 \sum_{m=1}^\infty \log (1+\CY_{m+1,1})\star 
\Big(H_m +H_{-m}\Big)   \\&\phaneq+ \sum_{m=1}^\infty \log (1+\CY_{m,0} ) \star \Sigma^{m} \,,
 \end{aligned}
\label{eq: rho}
\end{equation}
where 
\begin{equation}
 H_m(v,u) =\frac{i}{2\pi}\partial_v\log \frac{x(u-i0)-\frac{g^2}{x(v+\frac{i}{2}m)}}{x(u+i0)-\frac{g^2}{ x(v+\frac{i}{2}m)}}\,,
\end{equation}
\begin{equation}
\label{epsilon0}
\epsilon_0 (u) = \begin{cases}
                  0 &\mbox{for }  | u | \geq 2 g\,,\\
                  2\sqrt{4g^2-u^2} & \mbox{for }  |u| < 2g\,,
                 \end{cases}
\end{equation}
and the kernel
\begin{equation}
\begin{aligned}
\Sigma^{m} (v,u) =& \frac{i}{2\pi} \partial_v  \left( \log \frac{R^2(x(v+ \frac{im}{2}),x(u+i0))}{R^2(x(v+ \frac{im}{2}),x(u-i0))} \right. \\ &+ \left.  \log \frac{R^2(x(v- \frac{im}{2}),x(u-i0))}{R^2(x(v- \frac{im}{2}),x(u+i0))} \right) 
\end{aligned}
\end{equation}
is given in terms of the dressing factor \cite{Beisert:2006ez}
\begin{equation}
\begin{aligned}
\sigma^2 (u,v) =&  \frac{R^2(x^+(u),x^+(v))R^2(x^-(u),x^-(v))}{R^2(x^+(u),x^-(v))R^2(x^-(u),x^+(v))}      \,,
\end{aligned}
\end{equation}
with $x^\pm(u) = x(u\pm \frac{i}{2})$.
When applied to a function of two arguments such as $H_m(v,u)$, $\star$, $\hatstar$ and $\checkstar$ are moreover understood as integrals over the respective intervals.
The other exceptions to the  equations \eqref{eq: general Y system}
are the non-local equations
\begin{equation}
\label{eq: Y11_times_Y22}
\log \CY_{1,  1}\CY_{2,2} (u) = \sum_{m=1}^\infty \log (1+\CY_{m,0} (v) ) \star  \Theta_{m} (v,u) 
\end{equation}
with
\begin{equation}
\Theta_{m} (v,u) =\frac{i}{2\pi}\partial_v\log \frac{x(u)-\frac{g^2}{ x(v-\frac{i m}{2})}}{x(u)-\frac{g^2}{ x(v+\frac{i m}{2})}}\frac{x(u)-x(v+\frac{i m}{2})}{x(u)-x\left(v-\frac{i m}{2}\right)}
\end{equation}
and
\begin{equation}
\label{eq: Y11_divide_Y22}
\log \frac{\CY_{2,  2}}{\CY_{1,1}}  = \sum_{m=1}^\infty a_m \star \log \frac{ (1+ \CY_{m+1,  1} )^2}{(   1+ \CY_{1,m+1}^{-1} )^2(1+\CY_{m,0}  )} 
\end{equation}
with $a_n(u)=\frac{n}{2\pi(u^2+\frac{n^2}{4})}$.

The free energy per unit scaling dimension is given by 
\begin{equation}
\label{eq: free energy}
F(T)= - T \sum_{n=1}^\infty  \int_{-\infty}^\infty \de u \, \theta_n (u) \log ( 1 + \CY_{n,0} (u) )\eqncom
\end{equation}
where
\begin{equation}
\label{eq: theta n}
\theta_n(u) = \frac{i}{2\pi} \partial_u \log \frac{x(u+\frac{in}{2})}{x(u-\frac{in}{2})}\eqndot
\end{equation}
Thus, the TBA equations \eqref{eq: general Y system}--\eqref{eq: theta n} determine the Hagedorn temperature at any value of the 't Hooft coupling via \eqref{eq: hagedorn temperature via F}.

\section{Solving the TBA equations}

Let us now solve the TBA equations in the form of the Y-system.

\paragraph{Constant solution via T-system}

At large spectral parameter $u$, the Y-system approaches a constant value. This means we can find a constant Y-system that solves \eqref{eq: general Y system} for all $(a,s)\in M\setminus \{(1,1),(2,2)\}$ as well as \eqref{eq: Y11_divide_Y22}. 
Note that we cannot impose  \eqref{eq: Y11_times_Y22} as it relates the behavior at finite and large $u$. 
Thus, we find a one-parameter family of solutions with parameter $\noty$.
This solution is most easily expressed in terms of a T-system consisting of the functions $T_{a,s}$ with $(a,s)\in \hat{M}=\{(a,s)\in\mathbb{Z}_{\geq0}\times\mathbb{Z} \,|\,\min(a,|s|)\leq 2\}$ and $T_{a,s}=0$ for $(a,s)\notin \hat{M}$.
The Y-functions are expressed in terms of the T-functions as
\begin{equation}
 \CY_{a,s}=\frac{T_{a,s+1}T_{a,s-1}}{T_{a+1,s}T_{a-1,s}}\,.
\end{equation}
In the constant case, the  equations \eqref{eq: general Y system} imply the following T-system (Hirota) equations for all $(a,s)\in \hat{M}$:
\begin{equation}
\label{eq: T-system equations}
 T_{a,s}^2 = T_{a+1,s} T_{a-1,s} + T_{a,s+1} T_{a,s-1}\,.
\end{equation}
The latter are solved by
\begin{equation}
\label{consTbfgen1}
\begin{aligned}
T_{a,0} &=  \left(\frac{1-\yy}{1+\yy}\right)^{2a} \frac{a + 2\yy}{12 \yy^4}  \big( a^3 + 6 \yy a^2 \\
&\hphantom{=  \left(\frac{1-\yy}{1+\yy}\right)^{2a} \frac{a + 2\yy}{12 \yy^4}  \big(}+ ( 12\yy^2-1) a + 6\yy^3 \big) \,,
\\
T_{a,\pm 1} &=  (-1)^{a} \left(\frac{1-\yy}{1+\yy}\right)^{2a} \frac{a+3\yy}{6\yy^4}  (a^2+3a\yy + 3\yy^2-1) \,,
\\
T_{a,\pm 2} &= \frac{1}{\yy^4}  \left(\frac{1-\yy}{1+\yy}\right)^{2a}\,,
\end{aligned}
\end{equation}
for $a \geq |s|$, and
\begin{equation}
\label{consTbfgen2}
\begin{aligned}
T_{0,s}&=1\,,\\
T_{1,s} &= \frac{(-1)^s}{\yy^2} \left[ |s| + \frac{1-3\yy^2}{2\yy} \right] \left(\frac{1-\yy}{1+\yy}\right)^{|s|}  \,,\\
T_{2,s} &= \frac{1}{\yy^4}  \left(\frac{1-\yy}{1+\yy}\right)^{2|s|}\,,
\end{aligned}
\end{equation}
for $|s| \geq a$. 
This solution is a special case of the most general, $\mathfrak{psu}(2,2|4)$ character solution of \eqref{eq: T-system equations} in \cite{Gromov:2010vb}.

\paragraph{Solution at zero coupling}

In the limit of zero coupling, $g^2=0$, the source term $\rho(u)$ in \eqref{eq: CY_10} vanishes \footnote{In particular, $\text{constant}\star H_m=\text{constant}\star\Theta_m=0$.}, such that the functions $\CY_{a,s}$ are constant for all $u$. Hence, the non-local equation \eqref{eq: Y11_times_Y22} implies $\CY_{1,1}\CY_{2,2}=T_{1,0}=1$. We can use this to determine the parameter $\noty$ in the constant solution for the T-system above 
and thereby find the Y-system at zero coupling. Imposing $T_{1,0}=1$ is equivalent to $\noty = \pm 1/\sqrt{3}$. The negative solution has to be discarded as it leads to a negative Hagedorn temperature. Thus, we conclude that to zeroth order $\noty = 1/\sqrt{3}$. Using \eqref{eq: hagedorn temperature via F} and \eqref{eq: free energy}, we find the zeroth-order Hagedorn temperature 
\begin{equation}
 T_{\text{H}}^{(0)}=\frac{1}{2\log(2+\sqrt{3})}\,,
\end{equation}
which is in perfect agreement with \cite{Sundborg:1999ue}.

\paragraph{Perturbative solution}

We can also solve the TBA equations in a perturbative expansion at weak coupling, expanding the Y-functions as
\begin{equation}
\CY_{a,s} (u) = \CY_{a,s}^{(0)} \left( 1+ \sum_{\ell=1}^\infty g^{2\ell} y_{a,s}^{(\ell)} (u)\right) \,.
\end{equation}

At one-loop order, the solution takes the form
\begin{equation}
\label{eq: one-loop form}
y_{a,s}^{(1)}(u) =  \tilde{y}_{a,s}^{(1)}+  \sum_{k=0}^\infty c^{(1)}_{a,s,k} a_{2k+a+s}(u) \,,
\end{equation}
where $\tilde{y}_{a,s}^{(1)}$ as well as $c^{(1)}_{a,s,k}$ are constants.
This follows from the expansions
\begin{equation}
\label{eq: expansions of kernels}
 \begin{aligned}
   \epsilon_0\hatstar s(u)&=4\pi g^2 s(u) + 2\pi g^4 s''(u)+ \mathcal{O}(g^6)\,,\\
 s(u)&=\sum_{m=0}^\infty (-1)^{m}a_{1+2m}(u)\,, \\
   \theta_n(u)&=a_n(u)+g^2 a_n''(u)+\mathcal{O}(g^4)\,,\\
 \Theta_m(v,u)&=a_m(u-v)-a_m(v)\\&\phaneq+g^2\left(\frac{2}{u}a_m'(v)-a_m''(v)\right)+\mathcal{O}(g^4)
 \end{aligned}
\end{equation}
in combination with the convolution identity $a_n\star a_m=a_{n+m}$ and the structure of the TBA equations. Inserting \eqref{eq: one-loop form} into the expansion of the TBA equations, we can solve for the coefficients $c^{(1)}_{a,s,k}$.
The remaining one-loop parameter in the constant solution can be fixed from 
$\left(\CY_{1,1}\CY_{2,2}\right)^{(1)} (0)=0$, which follows from \eqref{eq: Y11_times_Y22} and the last expansion in \eqref{eq: expansions of kernels}.
We find for the one-loop Hagedorn temperature
\begin{equation}
 T_{\text{H}}^{(1)}=\frac{1}{\log(2+\sqrt{3})}\,,
\end{equation}
which perfectly agrees with the result of \cite{Spradlin:2004pp}.

At two-loop order, $\rho \hatstar s$ in \eqref{eq: CY_10} receives contributions from the one-loop solution
$y_{a,s}^{(1)}(u)$ from the second and third term in \eqref{eq: rho}. They can be calculated using 
\begin{equation}
 (a_n\star H_{m} \hatstar s)(u) = g^2 \frac{4}{(n+|m|)^2} s(u)  + \mathcal{O}(g^4)\,.
\end{equation}
Note that the dressing kernel in the fourth term of \eqref{eq: rho} vanishes at this loop order. The two-loop solution takes the form
\begin{equation}
\label{eq: two-loop form}
\begin{aligned}
y_{a,s}^{(2)}(u) &=  \tilde{y}_{a,s}^{(2)} + \sum_{k=0}^\infty c^{(2)}_{a,s,k,1} a_{2k+a+s}(u)\\ &\phaneq+  \sum_{k=0}^\infty c^{(2)}_{a,s,k,2} a^2_{2k+a+s}(u) \\ &\phaneq+  \sum_{k=0}^\infty c^{(2)}_{a,s,k,3} a^3_{2k+a+s}(u)   \,,
\end{aligned}
\end{equation}
as follows from simple reasoning paralleling the one at one-loop order.
Solving for the coefficients $c^{(2)}_{a,s,k,1}$, $c^{(2)}_{a,s,k,2}$ and $c^{(2)}_{a,s,k,3}$ and fixing the two-loop parameter in the constant solution via \eqref{eq: Y11_times_Y22}, we find the previously unknown two-loop Hagedorn temperature
\begin{equation}
\begin{aligned}
 T_{\text{H}}^{(2)}= 
  -\frac{86}{\sqrt{3}} +  \frac{24\log(12)}{\log(2+\sqrt{3})}\,.
\end{aligned}
\end{equation}

\paragraph{Solution at finite coupling}

At finite coupling, the infinite set of non-linear integral  equations \eqref{eq: general Y system}--\eqref{eq: theta n} can be solved numerically 
by iterating the equations and truncating to $a,s\leq n_{\max}$.
The convolutions are calculated for a finite number of sampling points from which the functions are recovered by interpolation and extrapolation at small and large $u$, respectively.
We have implemented this procedure in Mathematica following the strategy of \cite{Bajnok:2013wsa}, where also 
$T_{\text{H}}$ has to be iterated.
We will report on the resulting solution at finite coupling in our future publication \cite{HW}.

\section{Outlook}

In this letter, we have derived integrability-based TBA equations \eqref{eq: general Y system}--\eqref{eq: theta n} that determine the Hagedorn temperature of planar $\mathcal{N}=4$ SYM theory at any value of the 't Hooft coupling.
As an application, we have solved these equations perturbatively up to two-loop order.
Our TBA equation can also be solved numerically at finite coupling, as was briefly discussed here but will be detailed on in a future publication \cite{HW}.
Thus, they open up the door for an exact interpolation from weak to strong coupling, which, with the exception of \cite{Harmark:2006ta},  
would be the first time for the case of thermal physics.
Potentially, this could allow us to develop a better understanding of the phase structure of gauge theories and their dual gravitational theories in general.

For the spectral problem, the TBA equations have been recast into the form of the quantum spectral curve \cite{Gromov:2013pga}, which allows to generate precision data at weak coupling \cite{Marboe:2014gma} as well as at finite coupling \cite{Gromov:2015wca}.
We will report on a similar reformulation of our equations in a future publication \cite{HW}.
Moreover, one can study the case of non-zero chemical potentials. We have generalized our method to this case as well, and we have solved the zeroth-order TBA equations for the case with chemical potentials turned on but corresponding still to a symmetric Y-system. We will report on this in a future publication as well \cite{HW}.

In this letter, we have used the fact \eqref{eq: hagedorn temperature via F} that the spin-chain free energy determines the Hagedorn temperature $T_{\text{H}}$ at which the partition function diverges. The spin-chain free energy should however also determine the partition function in the vicinity of $T_{\text{H}}$, which should allow to extract e.g.\ critical exponents.

The partition function and Hagedorn temperature have also been studied in integrable deformations of $\mathcal{N}=4$ SYM theory up to one-loop order \cite{Fokken:2014moa}, where it was found that although $\mathcal{Z}(T)$ is different, $T_{\text{H}}$ is unchanged.
It would be interesting to see whether this statement continues to hold at higher loop orders. 
Similarly, one might apply our framework to the three-dimensional $\mathcal{N}=6$ superconformal Chern-Simons theory, which is known to be integrable as well.

\begin{acknowledgments}
\paragraph{Acknowledgements.}
It is a pleasure to thank Marta Orselli for collaboration in an earlier stage of the project.
We thank
Zoltan Bajnok,
Johannes Br\"{o}del,
Simon Caron-Huot,
Marius de Leeuw,
Sergey Frolov,
Nikolay Gromov,
Sebastien Leurent,
Fedor Levkovich-Maslyuk,
Christian Marboe,
David McGady,
Ryo Suzuki,
Dmytro Volin
and Konstantin Zarembo
for very useful discussions and 
Ryo Suzuki for sharing his Mathematica implementation of the TBA equations used in \cite{Bajnok:2013wsa}. 
T.H.\ acknowledges support from FNU grant number DFF-6108-00340 and the Marie-Curie-CIG grant number 618284.
M.W.\ was supported in part by FNU through grants number DFF-4002-00037 and by the ERC advance grant 291092.
M.W.\ further acknowledges the kind hospitality of NORDITA during the program ``Holography and Dualities,'' where parts of this work were carried out.
\end{acknowledgments}

\bibliography{mybib}

\end{document}